\documentclass[12pt]{iopart}
\makeatletter
\@namedef{ver@amsmath.sty}{}
\makeatother
\usepackage{xcolor}
\usepackage{graphicx}
\usepackage{bm}
\usepackage{amssymb}

\begin{document}

\title[Drift of suspended single-domain nanoparticles]{Drift of suspended single-domain nanoparticles in a harmonically oscillating gradient magnetic field}

\author{S I Denisov, T V Lyutyy and A T Liutyi}
\address{Sumy State University, Rimsky-Korsakov Street 2, UA-40007
Sumy, Ukraine}
\ead{denisov@sumdu.edu.ua}
\ead{lyutyy@oeph.sumdu.edu.ua}

\begin{abstract}
We study the nonlinear dynamics of single-domain ferromagnetic nanoparticles in a viscous liquid induced by a harmonically oscillating gradient magnetic field in the absence and presence of a static  uniform magnetic field. Under some physically reasonable assumptions, we derive a coupled set of stiff ordinary differential equations for the magnetization angle and particle coordinate describing the rotational and translational motions of nanoparticles. Analytical solutions of these equations are determined for nanoparticles near and far from the coordinate origin, and their correctness is confirmed numerically. We show that if a uniform magnetic field is absent, the magnetization angle and particle coordinate of each nanoparticle are periodic functions of time. In contrast, the presence of a uniform magnetic field makes these functions aperiodic. In this case, we perform a detailed analysis of the nanoparticle dynamics and predict the appearance of the drift motion (directed transport) of nanoparticles. We calculate both analytically and numerically the drift velocity, study its dependence on time and model parameters, analyze the physical origin of the drift phenomenon and discuss its potential biomedical applications.
\end{abstract}

\noindent{\it Keywords\/}: suspended single-domain nanoparticles, rigid dipole model, harmonically oscillating gradient magnetic field, coupled balance equations, drift phenomenon


\maketitle

\section{Introduction}
\label{intr}

Single-domain ferromagnetic nanoparticles in a viscous liquid possess interesting physical properties and promi\-sing applications in biomedicine \cite{PCJD2003, Ferr2005, CFR2018}. Many of these applications, including magnetic hyperthermia \cite{LDHM2011, PHS2015, DCP2019, R-R2021}, targeted drug delivery \cite{AFIS2007, SLZ2008, UHSB2016, LCSY2019} and cell separation \cite{ZC2008, PML2015, PS2020}, are mainly based on the rotational and translational properties of nano\-particles and their magnetization dynamics.

The rotation of suspended nanoparticles and their magnetization is usually caused by external and internal (thermal or dipolar) magnetic fields. In the context of magnetic hyperthermia, the rotational properties of nanoparticles with finite and infinite anisotropy (the latter case corresponds to the so-called rigid dipole mo\-del \cite{CKW2004}) driven by linearly and circularly polarized magnetic fields are studied in \cite{RS2011, UL2012, LR2018, LHY2019, URB2019}. Other important characteristics of suspended nanoparticles such as, e.g., induced magnetization, angular velocity, phase lag, nonexponential relaxation and dissipation-induced rotation, are considered in \cite{LDRB2015, Usad2017, IE2018, DLH2019}. Note that under some conditions ferromagnetic nanoparticles for magnetic hyperthermia can be considered immobilized \cite{PHS2015}. Therefore, results on the thermal stability of the magnetization precession modes induced by a rotating magnetic field \cite{DPL2011}, the energy dissipation in single-domain nanoparticles \cite{NR2012, LDPB2015, RCSSN2016} and the clasterization of nanoparticles because of their dipole-dipole interaction \cite{AZ2020} are also related to magnetic hyperthermia.

To be used in transport applications, it is necessary to control and manipulate the magnitude and direction of the driving force acting on suspended ferromagnetic nanoparticles. Often, to generate this force, the gradient magnetic fields, which can easily be produced by permanent or electromagnets, are used \cite{Svob2004}. These fields can also change the magnetization direction in nanoparticles and, due to their magnetic anisotropy, rotate them. One may therefore conclude that the gradient magnetic fields induce both the rotational and translational motions of suspended nanoparticles. These motions are, in general, coupled and their coupling can strongly affect the rotational and translational properties of such nanoparticles. Specifically, it is this coupling that is responsible for the existence of four regimes of directed transport of nanoparticles induced by the time-independent gradient and uniform magnetic fields \cite{DLP2020}.

In this paper, we study the coupled rotational and translational dynamics of suspended single-domain ferromagnetic nanoparticles driven by a harmonically oscillating gradient magnetic field in the presence of a static uniform magnetic field. Our main finding, which is confirmed by a detailed analytical and numerical analysis of the basic equations, is the existence of a novel mechanism of directed transport of such nanoparticles. We show that this mechanism is caused by a nonlinear coupling of the rotational and translational motions of nanoparticles and is realized through their drift in a preferred direction.

\section{Model and basic equations}
\label{sec:2}

In our model, the ferromagnetic nano\-par\-ticles are assumed to be spherical (with radius $a$), single-domain (i.e., their radius is less than the critical one \cite{Guim}) and noninteracting. If the anisotropy magnetic field is strong enough, then the direction of the domain magnetization $\mathbf{M} = \mathbf{M}(t)$ ($|\mathbf{M}| = M = \mathrm{const}$) can be chosen along the easy axis of anisotropy (the rigid dipole approximation). In this case, the magnetization dynamics is governed by the nanoparticle rotation and is described by the kinematic equation
\begin{equation}
    \frac{d}{dt}\mathbf{M} = \boldsymbol{
    \omega} \times \mathbf{M},
    \label{kinem}
\end{equation}
where $\boldsymbol{\omega} = \boldsymbol{\omega}(t)$ is the particle angular velocity and $\times$ denotes the vector product.

The harmonically oscillating gradient magnetic field $\mathbf{H}_{g}$ and the static uniform magnetic field $\mathbf{H}_{\perp}$ are chosen in the form
\begin{equation}
    \mathbf{H}_{g} = g x \sin{(\mathrm{
    \Omega} t + \phi)}\,\mathbf{e}_{x}, \quad
    \mathbf{H}_{\perp} = H_{\perp}
    \mathbf{e}_{y}.
    \label{H_p,H_g}
\end{equation}
Here, $g(>0)$, $\mathrm{\Omega}$ and $\phi \in [0, \pi]$ are, respectively, the gradient, angular frequency and initial phase of $\mathbf{H}_{g}$, $H_{\perp}(\geq 0)$ is the strength of $\mathbf{H}_{\perp}$ (the symbol $\perp$ means that $\mathbf{H}_{\perp}$ is perpendicular to $\mathbf{H}_{g}$), and $\mathbf{e}_{x}$ and $\mathbf{e}_{y}$ are the unit vectors along the $x$ and $y$ directions. The configuration of the total magnetic field $\mathbf{H}_{g} + \mathbf{H}_{\perp}$ allows to represent the nanoparticle magnetization in the form
\begin{equation}
    \mathbf{M} = M (\cos{\varphi}\,\mathbf{e}_{x}
    + \sin{\varphi}\,\mathbf{e}_{y})
    \label{M}
\end{equation}
with $\varphi = \varphi(t) \in [0, \pi]$ being the magnetization angle.

In the inertialess approximation, the dynamical eq\-uations describing the rotational and translational motions of nanoparticles are reduced to the balance torque and force equations, $\mathbf{t}_{d} + \mathbf{t}_{f} = 0$ and $\mathbf{f}_{d} + \mathbf{f}_{f} = 0$, respectively. The driving torque $\mathbf{t}_{d}$ and force $\mathbf{f}_{d}$ are defined in the standard way: $\mathbf{t}_{d} = V \mathbf{M} \times (\mathbf{H}_{g} + \mathbf{H} _{\perp})|_{x=R_{x}}$ and $\mathbf{f}_{d} = V(\mathbf{M} \cdot \nabla) \mathbf{H}_{g}|_{x=R_{x}}$, where $V=4\pi a^{3}/3$ is the particle volume, $\nabla$ is the vector differential operator del, the symbol $\cdot$ denotes the scalar product, and $R_{x} = R_{x}(t)$ is the $x$-component of the particle centre. Using (\textcolor{blue}{\ref{H_p,H_g}}) and (\textcolor{blue}{\ref{M}}), one can easily show that $\mathbf{t}_{d} = t_{d} \mathbf{e}_{z}$ ($\mathbf{e}_{z}$ is the unit vector along the $z$ direction) with
\begin{equation}
    t_{d} = MV [H_{\perp} \cos{\varphi} -
    gR_{x} \sin{\varphi} \sin{(\mathrm{
    \Omega} t + \phi)}],
    \label{t_d}
\end{equation}
and $\mathbf{f}_{d} = f_{d} \mathbf{e}_{x}$ with
\begin{equation}
    f_{d} = MVg \cos{\varphi} \sin{(\mathrm{
    \Omega} t + \phi)}.
    \label{f_g}
\end{equation}
Since in our approach the rotational and translational Rey\-nolds numbers are small, for the frictional torque $\mathbf{t}_{f}$ and force $\mathbf{f}_{f}$ we use the expressions $\mathbf{t}_{f} = -6\eta V \boldsymbol{\omega}$ and $\mathbf{f}_{f} = -6\pi \eta a \mathbf{V}$ derived by Kirchhoff and Stokes (these expressions represent the leading terms in the expansions of $\mathbf{t}_{f}$ and $\mathbf{f}_{f}$, see, e.g., \cite{RuKe1961}). Here, $\eta$ is the dynamic viscosity of liquid and $\mathbf{V} = (dR_{x}/dt)\mathbf{e}_{x}$ is the particle translational velocity.

Within these approximations, the balance tor\-que equation yields $\boldsymbol{\omega} = \omega_{z} \mathbf{e}_{z}$, where
\begin{equation}
    \omega_{z} = \frac{M}{6\eta} [H_{\perp}
    \cos{\varphi} -  gR_{x} \sin{\varphi}
    \sin{(\mathrm{\Omega} t + \phi)}],
    \label{omega_z}
\end{equation}
and the balance force equation reduces to
\begin{equation}
    \frac{d}{dt}R_{x} = \frac{2Mga^{2}}
    {9\eta} \cos{\varphi} \sin{(\mathrm{
    \Omega} t + \phi)}.
    \label{V_x}
\end{equation}
Finally, using the relation $\omega_{z} = d\varphi /dt$, which follows from (\textcolor{blue}{\ref{kinem}}) and (\textcolor{blue}{\ref{M}}), and introducing the dimensionless time $\tau = \mathrm{\Omega} t$ and particle coordinate $r_{x} = R_{x}/a$, (\textcolor{blue}{\ref{omega_z}}) and (\textcolor{blue}{\ref{V_x}}) can be rewritten in the dimensionless form as
\begin{equation}
    \dot{\varphi} = \nu_{\perp}
    \cos{\varphi} - \nu_{g}r_{x}\sin{\varphi}
    \sin{(\tau + \phi)}
    \label{set_1}
\end{equation}
and
\begin{equation}
    \dot{r}_{x} = \case{4}{3}\nu_{g}
    \cos{\varphi} \sin{(\tau + \phi)},
    \label{set_2}
\end{equation}
respectively. Here, the overdot denotes the derivative with respect to $\tau$ and
\begin{equation}
    \nu_{\perp} = \frac{MH_{\perp}}
    {6\eta \mathrm{\Omega}}, \quad
    \nu_{g} = \frac{Mga}{6\eta\mathrm{
    \Omega}}
    \label{nu}
\end{equation}
are the dimensionless characteristic frequencies arising from the uniform and gradient magnetic fields. We note that equations (\textcolor{blue}{\ref{set_1}}) and (\textcolor{blue}{\ref{set_2}}) generalize the corresponding equations derived in \cite{DLP2020} for a time-independent gradient magnetic field.

Equations (\textcolor{blue} {\ref{set_1}}) and (\textcolor{blue} {\ref{set_2}}), together with the initial conditions $\varphi(0) = \varphi_{0} \in [0, \pi]$ and $r_{x}(0) = r_{x0} \in (-\infty, \infty)$, describe the coupled dynamics of the variables $\varphi$ and $r_{x}$. According to them, the harmonically oscillating gradient magnetic field directly affects the rotational and translational motions of ferromagnetic nanoparticles. In contrast, the uniform magnetic field directly affects only their rotational motion. Nevertheless, due to the coupling between the magnetization angle and the particle coordinate, this field qualitatively changes the translational dynamics of nanoparticles (see section \textcolor{blue} {\ref{sec:4}}).

Note also that the set of coupled first-order differential equations (\textcolor{blue} {\ref{set_1}}) and (\textcolor{blue}{\ref{set_2}}) can be reduced to two uncoupled equations for $\varphi$ and $r_{x}$. Indeed, calculating the derivative $\dot{r}_{x}$ from (\textcolor{blue} {\ref{set_1}}) and substituting it into (\textcolor{blue} {\ref{set_2}}), one obtains the following second-order differential equation for the magnetization angle:
\begin{eqnarray}
    &\ddot{\varphi}\sin{\varphi} + \dot{\varphi}
    \left[\nu_{\perp} - \sin{\varphi}\cot{(\tau +
    \phi)}\right]\nonumber \\[3pt]
    &- \dot{\varphi}^{2} \cos{\varphi} +
    \nu_{\perp} \sin{\varphi} \cos{\varphi}
    \cot{(\tau + \phi)}\nonumber \\[3pt]
    &+ \case{4}{3} \nu_{g}^{2} \sin^{2}{\varphi}
    \cos{\varphi} \sin^{2} {(\tau + \phi)} =0,
    \label{eq_varphi}
\end{eqnarray}
whose solution must satisfy the initial conditions
\begin{equation}
    \varphi(0) = \varphi_{0}, \quad
    \dot{\varphi}(0) = \nu_{\perp}\cos{\varphi_{0}}
    - \nu_{g} r_{x0} \sin{\varphi_{0}}\sin{\phi}.
    \label{init1}
\end{equation}
Similarly, it can be shown that the dimensionless particle coordinate is described by the second-order differential equation
\begin{eqnarray}
    &\ddot{r}_{x} + \dot{r}_{x}[\nu_{\perp}
    \sin{\varphi} - \cot{(\tau + \phi)}]
    \nonumber \\[3pt]
    &- \case{4}{3} \nu_{g}^{2} r_{x}\sin^{2}{\varphi}
    \sin^{2} {(\tau + \phi)} =0,
    \label{eq_r_x}
\end{eqnarray}
where
\begin{equation}
    \sin{\varphi} = \left[1- \left( \frac{3 \dot{r}_{x}}
    {4 \nu_{g} \sin{(\tau + \phi)}} \right)^{2}
    \right]^{1/2},
    \label{rel1}
\end{equation}
with the initial conditions
\begin{equation}
    r_{x}(0) = r_{x0}, \quad
    \dot{r}_{x}(0) = \case{4}{3}\nu_{g}
    \cos{\varphi_{0}} \sin{\phi}.
    \label{init2}
\end{equation}

It should also be emphasized that the above equations of motion describe the rotational and translational motions of noninteracting nanoparticles. Therefore these equations are applicable only if the average distance $l$ between nanoparticles greatly exceeds their diameter: $l\gg d$. Since the concentration $c$ of nanoparticles is of the order of $l^{-3}$, the last condition can be rewritten as $c \ll d^{-3}$.

\section{Nanoparticle dynamics at $\boldsymbol{\nu_{\perp} = 0}$}
\label{sec:3}

Before starting our analysis, we note that equations (\textcolor{blue} {\ref{set_1}}) and (\textcolor{blue} {\ref{set_2}}) with $\nu_{\perp}=0$ are a particular case of equations used by Massera solely to illustrate his theorem on the existence of periodic solutions of two-dimensional ordinary differential equations with periodic right-hand sides \cite{Mass1950}. But here we solve these equations both analytically and numerically under different conditions. Moreover, while the illustrative equations had no physical meaning (they were introduced for illustrative purposes only), each of equations (\textcolor{blue} {\ref{set_1}}) and (\textcolor{blue} {\ref{set_2}}) has a clear physical interpretation.

\subsection{Particular analytical solutions}
\label{ssec:3.1}

The set of equations (\textcolor{blue} {\ref{set_1}}) and (\textcolor{blue} {\ref{set_2}}) can easily be solved for $\varphi_{0}=0$ and $\varphi_{0} =\pi$. By direct integration, we find for the first and second cases
\begin{eqnarray}
    &\varphi=0, \quad
    r_{x} = r_{x0} + \case{4}{3}\nu_{g}f(\tau),
    \nonumber \\[3pt]
    &\varphi=\pi, \quad
    r_{x} = r_{x0} - \case{4}{3}\nu_{g}f(\tau),
    \label{sol_0,pi}
\end{eqnarray}
respectively, where
\begin{equation}
    f(\tau) = \cos{\phi} - \cos{(\tau + \phi)}.
    \label{def_f}
\end{equation}
In agreement with physical expectations, the time-vary\-ing gradient magnetic field at $\varphi_{0}=0, \pi$ does not cause the nanoparticle rotation; it only causes their translational oscillations. Note, the same solutions (\textcolor{blue} {\ref{sol_0,pi}}) can also be obtained from equations (\textcolor{blue} {\ref{eq_varphi}}) and (\textcolor{blue} {\ref{eq_r_x}}).

\subsection{Exact results}
\label{ssec:3.2}

For $\varphi_{0} \neq 0, \pi$, we were not able to solve equations (\textcolor{blue} {\ref{set_1}}) and (\textcolor{blue} {\ref{set_2}}) exactly. However, in this case it is possible to establish some exact relations for $\varphi$ and $r_{x}$. In order to derive them, we first rewrite these equations as
\begin{equation}
    \frac{d}{d\tau}\ln{\sin{\varphi}} =
    - \nu_{g}r_{x}\cos{\varphi}
    \sin{(\tau + \phi)}
    \label{set_12}
\end{equation}
and
\begin{equation}
    \frac{d}{d\tau}r_{x}^{2} = \frac{8}{3}
    \nu_{g}r_{x}\cos{\varphi} \sin{(\tau + \phi)}.
    \label{set_22}
\end{equation}
Then, dividing (\textcolor{blue} {\ref{set_12}}) by (\textcolor{blue} {\ref{set_22}}), one obtains the equation
\begin{equation}
    \frac{d}{d\tau}\left( \ln{\sin{\varphi}}
    + \frac{3}{8} r_{x}^{2} \right) = 0
    \label{diff_rel}
\end{equation}
that after integration over $\tau$ from 0 to $\tau$ leads to the algebraic relation
\begin{equation}
    \ln\frac{\sin{\varphi}}{\sin{
    \varphi_{0}}} + \frac{3}{8}(r_{x}^{2}
    - r_{x0}^{2}) = 0,
    \label{rel_1}
\end{equation}
which will be used below.

The second relation we obtain dividing equation (\textcolor{blue} {\ref{set_1}}) by $\sin{\varphi}$ and integrating its both sides from 0 to $\tau$. Using the standard integral
\begin{equation}
    \int \frac{dx}{\sin{x}} = \frac{1}{2}
    \ln{\frac{1-\cos{x}}{1+\cos{x}}}
    \label{int_1}
\end{equation}
and introducing the function
\begin{equation}
    Y(\tau) = \nu_{g} \int_{0}^{\tau}r_{x}(\tau')
    \sin{(\tau' + \phi)} d\tau',
    \label{def_Y}
\end{equation}
this yields
\begin{equation}
    \cos{\varphi} = \frac{\cos{\varphi_{0}}
    + \tanh{Y(\tau)}} {1 + \cos{\varphi_{0}}
    \tanh{Y(\tau)}}.
    \label{cos1}
\end{equation}
Next we multiply equation (\textcolor{blue} {\ref{set_2}}) by $r_{x}$, which after substitution of the previous result and integration from 0 to $\tau$ gives
\begin{equation}
    r_{x}^{2} - r_{x0}^{2} = \frac{8}{3}
    \int_{0}^{Y(\tau)} \frac{\cos{\varphi_{0}}
    + \tanh{Y}} {1 + \cos{\varphi_{0}}
    \tanh{Y}} dY.
    \label{rel_1_2}
\end{equation}
Finally, using the standard integral
\begin{equation}
    \int \frac{\cos{\varphi_{0}} + \tanh{x}}
    {1 + \cos{\varphi_{0}} \tanh{x}} dx =
    \ln{\cosh{\left(x - \ln{\tan{\frac{
    \varphi_{0}}{2}}}  \right)}},
    \label{int_2}
\end{equation}
formula (\textcolor{blue} {\ref{rel_1}}) can be reduced to
\begin{equation}
    r_{x}^{2} - r_{x0}^{2} = \frac{8}{3}
    \ln{\frac{\cosh{\left(Y(\tau) - \ln{\tan{
    \frac{\varphi_{0}}{2}}} \right)}}
    {\cosh{\left(\ln{\tan{\frac{
    \varphi_{0}}{2}}}\right)}}}.
    \label{rel_1_3}
\end{equation}

A simple analysis shows that the asymptotic growth of the function $|Y(\tau)|$ and the right side of (\textcolor{blue} {\ref{rel_1_3}}) cannot be faster than $\tau$. On the other hand, if the drift motion of nanoparticles exists, than $r_{x}^{2}(\tau) \sim \tau^{2}$ as $\tau \to \infty$ (see section \textcolor{blue} {\ref{sec:4}}). This contradiction in the asymptotic behavior of the right and left sides of (\textcolor{blue} {\ref{rel_1_3}}) suggests that  that there is no the drift motion if $\nu_{\perp}=0$.

\subsection{Approximate solutions}
\label{ssec:3.3}

Now we solve equations (\textcolor{blue} {\ref{set_1}}) and (\textcolor{blue} {\ref{set_2}}) approximately assuming that $|r_{x} - r_{x0}| \ll 1$. In this case, replacing in (\textcolor{blue} {\ref{set_1}}) $r_{x}$ by $r_{x0}$ and using the integral (\textcolor{blue} {\ref{int_1}}), one obtains the following expression for the time-dependent magnetization angle:
\begin{equation}
    \varphi = \arccos{\left(\frac{\cos{\varphi_{0}}
    + \tanh{[\nu_{g}r_{x0}f(\tau)]}}{1 +
    \cos{\varphi_{0}} \tanh{[\nu_{g}r_{x0}
    f(\tau)]}}\right)}.
    \label{varphi_1}
\end{equation}
Then, substituting (\textcolor{blue} {\ref{varphi_1}}) into (\textcolor{blue} {\ref{rel_1}}) and taking into account that in our approximation $r_{x}^{2} - r_{x0}^{2} = 2r_{x0}(r_{x} - r_{x0})$ (if $r_{x0} \neq 0$), after some simple algebra we find the time evolution of the particle coordinate
\begin{equation}
    r_{x} = r_{x0} + \case{4}{3}\nu_{g}
    \ln{[G(\tau)]^{\frac{1}{\nu_{g} r_{x0}}}},
    \label{r_x_1}
\end{equation}
where
\begin{equation}
    G(\tau) = \cosh{[\nu_{g} r_{x0} f(\tau)]}
    + \cos{\varphi_{0}} \sinh{[\nu_{g}
    r_{x0} f(\tau)]}.
    \label{G(tau)}
\end{equation}

It should be noted that, according to (\textcolor{blue} {\ref{def_f}}), the magnetization angle (\textcolor{blue} {\ref{varphi_1}}) is periodic with the dimensionless period $2\pi$. In general, the particle coordinate $r_{x}$ is also periodic with the same period. But, as it follows from (\textcolor{blue} {\ref{def_f}}), (\textcolor{blue} {\ref{r_x_1}}) and (\textcolor{blue} {\ref{G(tau)}}), in the special case when the conditions $\phi = \pi/2$ and $\varphi_{0} = \pi/2$ are satisfied simultaneously, the period of $r_{x}$ equals $\pi$.

The functions (\textcolor{blue} {\ref{varphi_1}}) and (\textcolor{blue} {\ref{r_x_1}}) represent the solution of equations (\textcolor{blue} {\ref{set_1}}) and (\textcolor{blue} {\ref{set_2}}) at $|r_{x} - r_{x0}| \ll 1$. Since the maximum value of $[G(\tau)]^{\frac{1}{\nu_{g} r_{x0}}}$ is of the order of a few units, from (\textcolor{blue} {\ref{r_x_1}}) it follows that the condition $|r_{x} - r_{x0}| \ll 1$ holds only if $\nu_{g} \ll 1$. It is important to emphasize that the values of $|r_{x0}|$ can be arbitrary, i.e., the rotational and translational motions of all particles at $\nu_{g} \ll 1$ are described by the same formulas (\textcolor{blue} {\ref{varphi_1}}) and (\textcolor{blue} {\ref{r_x_1}}). In particular, for particles near the origin, when $\nu_{g} |r_{x0}| \ll 1$, they can be simplified to
\begin{eqnarray}
    &\varphi= \varphi_{0} - \nu_{g}r_{x0}
    \sin{\varphi_{0}}f(\tau) - \case{1}{4}
    \nu_{g}^{2}r_{x0}^{2} \sin{(2\varphi_{0})}
    f^{2}(\tau), \nonumber \\[3pt]
    &r_{x} = r_{x0} + \case{4}{3}\nu_{g}
    \cos{\varphi_{0}}f(\tau) + \case{2}{3}
    \nu_{g}^{2}r_{x0} \sin^{2}{\varphi_{0}}
    f^{2}(\tau).
    \label{varphi,r_x_2}
\end{eqnarray}
These results, obtained in the second-order approximation in $\nu_{g}$ and $\nu_{g} |r_{x0}|$, at $\varphi_{0} = 0,\pi$ are confirmed by the exact solutions (\textcolor{blue} {\ref{sol_0,pi}}). Note that, according to (\textcolor{blue} {\ref{varphi,r_x_2}}), the amplitude of the magnetization angle oscillations grows much more rapidly with increasing $|r_{x0}|$ than the amplitude of the coordinate oscillations.

For nanoparticles that are very far away from the origin, i.e., when $\nu_{g}|r_{x0}| \gg 1$, formula (\textcolor{blue} {\ref{varphi_1}}) asymptotically reduces to the $2\pi$-periodic step function
\begin{equation}
    \varphi = \left\{\!\! \begin{array}{ll}
    0, & \; r_{x0} f(\tau)>0,
    \\[3pt]
    \varphi_{0}, & \; f(\tau)=0,
    \\[3pt]
    \pi, & \; r_{x0} f(\tau)<0.
    \end{array}
    \right.
\label{varphi_2}
\end{equation}
In general, the time dependence of the particle coordinate in this case can be determined from (\textcolor{blue} {\ref{r_x_1}}) and (\textcolor{blue} {\ref{G(tau)}}) as well. However, for this purpose, it is more convenient to use the integral form of equation (\textcolor{blue} {\ref{set_2}})
\begin{equation}
    r_{x} = r_{x0} + \frac{4}{3} \nu_{g}
    \int_{0}^{\tau} \cos{\varphi(\tau')}
    \sin{(\tau' + \phi)}d\tau'.
    \label{r_x}
\end{equation}
Indeed, substituting (\textcolor{blue} {\ref{varphi_2}}) into (\textcolor{blue} {\ref{r_x}}) and using the conditions that $f(\tau)>0$ [$f(\tau)<0$] at $\tau \in (0, 2\pi - 2\phi)$ and $f(\tau)<0$ [$f(\tau)>0$] at $\tau \in (2\pi - 2\phi, 2\pi)$ if $\phi \in (0, \pi/2)$ [$\phi \in (\pi/2, \pi)$], from (\textcolor{blue} {\ref{r_x}}) one straightforwardly obtains
\begin{equation}
    r_{x} = r_{x0} \pm \case{4}{3}
    \nu_{g} |f(\tau)|,
    \label{r_x_2}
\end{equation}
where the upper and lower signs correspond to $r_{x0} > 0$ and $r_{x0} < 0$, respectively. Thus, in the reference case, the particle oscillations are located just to the right (left) of the point $r_{x0}$ if $r_{x0} > 0$ ($r_{x0} < 0$). It should also be emphasized that the period of these oscillations abruptly decreases from $2\pi$ to $\pi$ if $\phi = \pi/2$, i.e., the second condition $\varphi_{0} = \pi/2$ is not necessary in the limit $\nu_{g}|r_{x0}| \to \infty$. We note that the nanoparticle dynamics at $\nu_{g} \sim 1$ and $|r_{x0}| \gg 1$ is described by the same formulas (\textcolor{blue} {\ref{varphi_2}}) and (\textcolor{blue} {\ref{r_x_2}}).

\subsection{Numerical verification}
\label{ssec:3.4}

To gain more insight into the coupled rotational and translational dynamics of nanoparticles and to verify the above analytical results, we solved equations (\textcolor{blue} {\ref{set_1}}) and (\textcolor{blue} {\ref{set_2}}) with $\nu_{\perp} = 0$ numerically. Our numerical results are presented for $\mathrm{SmCo}_{5}$ nanoparticles moving in water at room temperature ($295\, \mathrm{K}$) and characterized by the parameters $M = 1.36 \times 10^{3}\, \mathrm{emu\, cm^{-3}}$ and $\eta = 9.62\times 10^{-3}\, \mathrm{P}$. The choice of $\mathrm{SmCo}_{5}$ nanoparticles is mainly motivated by two reasons. First, in this material the energy of magnetic anisotropy is so large that the approximation of `frozen' magnetization is well justified. And second, since the critical diameter $2a_{cr}$ for $\mathrm{SmCo}_{5}$ nanoparticles is of the order of $10^{3}\, \mathrm{nm}$ \cite{Guim}, the single-domain condition $a < a_{cr}$ holds even for large nanoparticles (we choose $a = 10^{2}\, \mathrm{nm}$). The gradient $g$ of the harmonically oscillating gradient magnetic field is chosen to be $10^{2}\, \mathrm{Oe\, cm^{-1}}$ and the angular frequency $\mathrm{\Omega}$ of this field is assumed to be such that the dimensionless characteristic frequency $\nu_{g}$ ranges from $10^{-3}$ ($\mathrm{\Omega} \approx 2.36 \times 10^{4}\,\mathrm{rad\, s^{-1}}$) to $10^{-1}$ ($\mathrm{\Omega} \approx 2.36 \times 10^{2}\,\mathrm{rad\, s^{-1}}$). Further, in accordance with our previous assumptions, we choose the phase $\phi$ of the gradient magnetic field and the initial magnetization angle $\varphi_{0}$ from the interval $[0, \pi]$. At the same time, the possible values of the inial particle position $r_{x0}$ are restricted by the system size. Therefore, assuming that the suspension is confined between parallel planes $x=\pm l$ ($l = 1 \mathrm{cm}$), one obtains $|r_{x0}| \leq l/a = 10^{5}$.

The fact that the condition $\nu_{g} |r_{x0}| \gg 1$ can be satisfied
causes some technical difficulties in solving the set of equations (\textcolor{blue} {\ref{set_1}}) and (\textcolor{blue} {\ref{set_2}}) numerically. The reason is that the nano\-particle system in this case is characterized by two very different (dimensionless) time scales $\tau_{1}$ and $\tau_{2}$. The first one characterizes the translational dynamics of nanoparticles under the harmonically oscillating gradient magnetic field; according to (\textcolor{blue} {\ref{r_x_2}}) it can be defined as $\tau_{1} = 1$. And the second one characterizes the time interval during which the magnetization angle changes from $0$ to $\pi$ (or vice versa). Using (\textcolor{blue} {\ref{varphi_1}}) and (\textcolor{blue} {\ref{def_f}}), this time interval can be estimated as $\tau_{2} = (\nu_{g}|r_{x0}|)^{-1}$, i.e., $\tau_{2} \ll \tau_{1}$. Therefore, since the time step $\Delta \tau$ must satisfy the condition $\Delta \tau \ll \mathrm{min}\! \left\{\tau_{1}, \tau_{2}\right\}$, its size can be so small that the numerical solution of equations (\textcolor{blue} {\ref{set_1}}) and (\textcolor{blue} {\ref{set_2}}) becomes impractical for $\nu_{g} |r_{x0}| \gg 1$. This problem is common to all stiff ordinary differential equations, and a number of methods have been developed for its solution, see, e.g., \cite{HaWa2002}.

\begin{figure}
    \centering
    \includegraphics[totalheight=5.5cm]{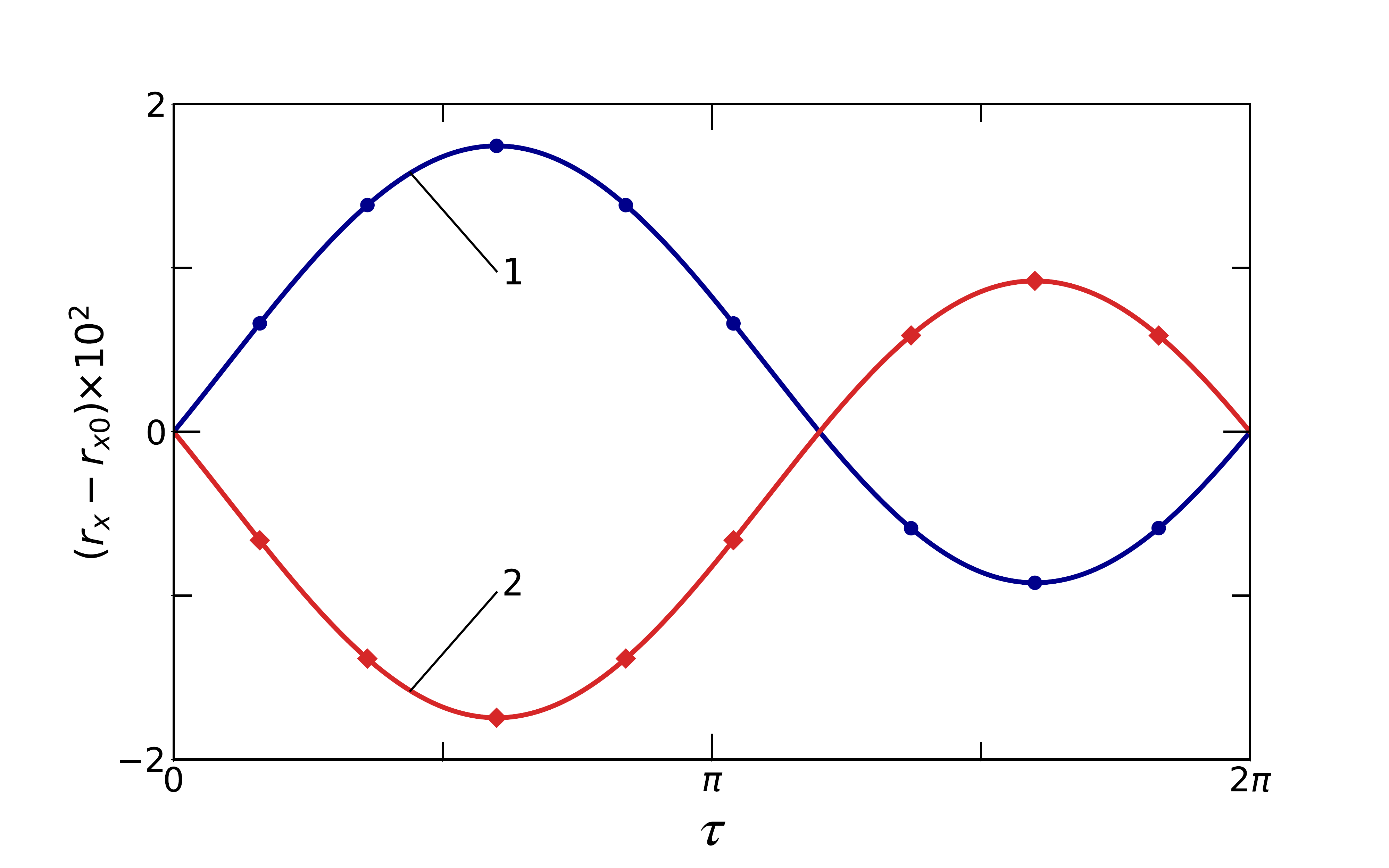}
    \caption{Particle coordinate $r_{x}$ as a function
    of the dimensionless time $\tau$ for $\varphi_{0} =
    0$ (line 1) and $\varphi_{0} = \pi$ (line 2). The
    other parameters are chosen as $\nu_{g} = 10^{-2}$,
    $\phi = 0.4\pi$ and the initial particle
    position $r_{x0}$ is arbitrary. The solid
    lines represent exact results from
    (\textcolor{blue} {\ref{sol_0,pi}}), while
    the results obtained from the numerical
    solution of equations (\textcolor{blue} {\ref{set_1}}) and
    (\textcolor{blue} {\ref{set_2}}) are shown
    by symbols.}
    \label{fig1}
\end{figure}

\begin{figure}
    \centering
    \includegraphics[totalheight=5.5cm]{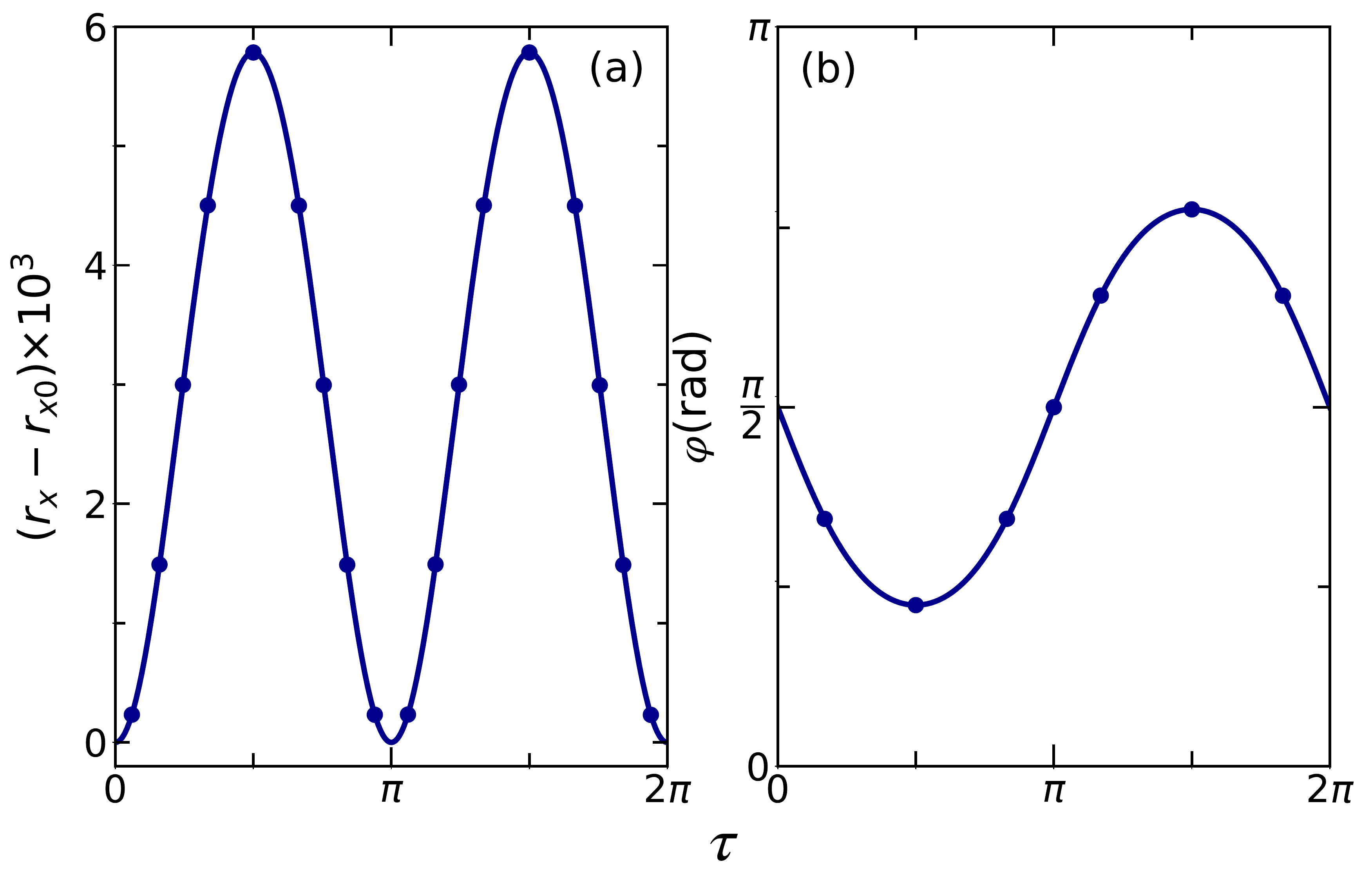}
    \caption{Dependence of the particle coordinate
    $r_{x}$ (a) and the magnetization angle $\varphi$
    (b) on the dimensionless time $\tau$ for $\nu_{g}
    = 10^{-2}$, $r_{x0} = 10^{2}$, $\varphi_{0} = \pi/2$
    and $\phi = \pi/2$. The solid lines correspond to
    the approximate solution, (\textcolor{blue} {\ref{r_x_1}}) and
    (\textcolor{blue} {\ref{varphi_1}}), of equations (\textcolor{blue} {\ref{set_1}}) and
    (\textcolor{blue} {\ref{set_2}}) for
    $r_{x}$ and $\varphi$, respectively. As in
    figure \textcolor{blue} {\ref{fig1}}, the results of the numerical
    solution of these equations
    are shown by symbols.}
    \label{fig2}
\end{figure}

\begin{figure}
    \centering
    \includegraphics[totalheight=5.5cm]{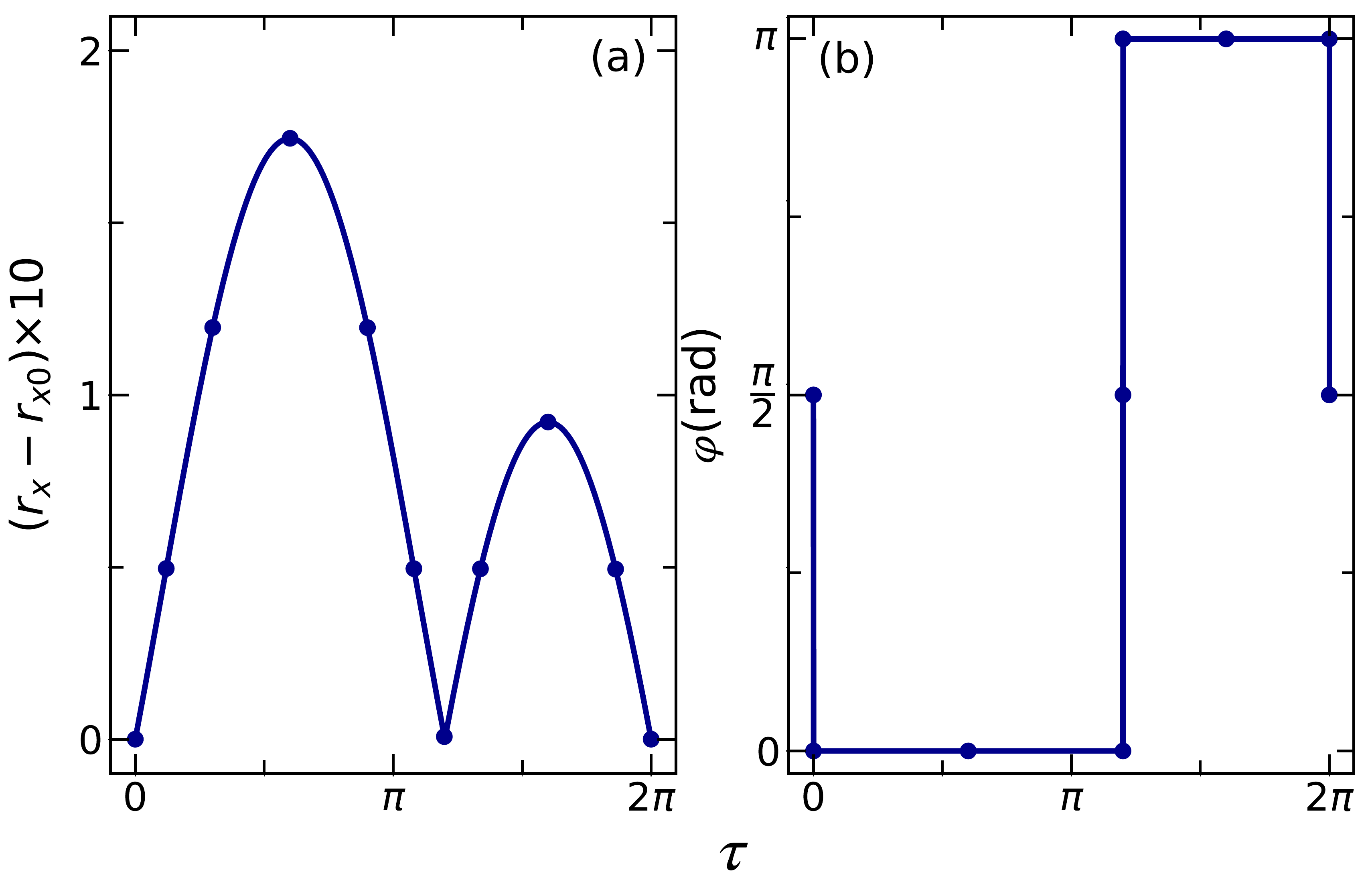}
    \caption{Dependence of the particle coordinate
    $r_{x}$ (a) and the magnetization angle $\varphi$
    (b) on the dimensionless time $\tau$ for $\nu_{g} =
    10^{-1}$, $r_{x0} = 10^{4}$, $\varphi_{0} = \pi/2$
    and $\phi = 0.4\pi$. The functions (\textcolor{blue} {\ref{r_x_2}})
    and (\textcolor{blue} {\ref{varphi_2}}), which represent the
    approximate solution of equations (\textcolor{blue} {\ref{set_1}}) and
    (\textcolor{blue} {\ref{set_2}}) for $r_{x}$ and $\varphi$,
    respectively, are shown by the solid lines.
    The symbols have the same meaning as before.}
    \label{fig3}
\end{figure}

Using the Runge–Kutta method of the fourth order, we solved equations (\textcolor{blue} {\ref{set_1}}) and (\textcolor{blue} {\ref{set_2}}) and confirmed our theoretical results. In figure \textcolor{blue} {\ref{fig1}}, we show that the analytical solutions (\textcolor{blue} {\ref{sol_0,pi}}) of these equations are in agreement with the numerical ones.
Figure \textcolor{blue} {\ref{fig2}} illustrates the correctness of the approximate solution, (\textcolor{blue} {\ref{varphi_1}}) and (\textcolor{blue} {\ref{r_x_1}}), of equations (\textcolor{blue} {\ref{set_1}}) and (\textcolor{blue} {\ref{set_2}}) at $\nu_{g} \ll 1$. It also confirms our prediction that the period of the particle coordinate $r_{x}$ at $\varphi_{0} = \phi = \pi/2$ is two times less than the period of the magnetization angle $\varphi$. Finally, the case with $\nu_{g} |r_{x0}| \gg 1$ is illustrated in figure \textcolor{blue} {\ref{fig3}}. As seen, the approximate and numerical solutions of (\textcolor{blue} {\ref{set_1}}) and (\textcolor{blue} {\ref{set_2}}) are indistinguishable as well.
Note that, according to (\textcolor{blue} {\ref{varphi_2}}) and (\textcolor{blue} {\ref{r_x_2}}), further increasing $|r_{x0}|$ does not affect the time dependence of $r_{x} - r_{x0}$ and $\varphi$, but the numerical solution of these equations becomes very computationally expensive.

Thus, our analytical and numerical results unambiguously show that if $\nu_{\perp} = 0$, then the harmonically oscillating gradient magnetic field induces only periodic motions (translational and rotational) of nanoparticles.

\section{Nanoparticle dynamics at $\boldsymbol{\nu_{\perp} > 0}$}
\label{sec:4}

\subsection{Some preliminary results and definitions}
\label{ssec:4.1}

We start the analysis of the nanoparticle dynamics, which occurs at $\nu_{\perp} \neq 0$, with the generalization of the relation (\textcolor{blue} {\ref{rel_1}}). Proceeding in the same manner as in section \textcolor{blue} {\ref{sec:2}} and taking into account that
\begin{equation}
    \frac{d}{d\tau}\left( \ln{\sin{\varphi}}
    + \frac{3}{8} r_{x}^{2} \right) =
    \nu_{\perp} \frac{\cos^{2}{\varphi}}
    {\sin{\varphi}},
    \label{diff_rel2}
\end{equation}
we obtain the integral relation
\begin{equation}
    \ln\frac{\sin{\varphi}}{\sin{
    \varphi_{0}}} + \frac{3}{8}(r_{x}^{2}
    - r_{x0}^{2}) = \nu_{\perp}
    \int_{0}^{\tau} \frac{\cos^{2}{\varphi
    (\tau')}}{\sin{\varphi(\tau')}} d\tau',
    \label{rel_2}
\end{equation}
which generalizes (\textcolor{blue} {\ref{rel_1}}) to the case when the static magnetic field $H_{\perp}$ is non-zero ($\nu_{\perp} \neq 0$). Since $\varphi \in [0, \pi]$, its right-hand side, $P(\tau)$, monotonically grows with time. Generally speaking, there are two regimes of growth. The first one is characterized by the condition $P(\infty) < \infty$, which, according to (\textcolor{blue} {\ref{set_1}}) and (\textcolor{blue} {\ref{set_2}}), can be realized if $\varphi \to \pi/2$ and $r_{x} \to 0$ as $\tau \to \infty$. As follows from (\textcolor{blue} {\ref{rel_2}}), this regime may exist only for nanoparticles, whose initial magnetization angle $\varphi_{0}$ and position $r_{x0}$ satisfy the condition
\begin{equation}
    \ln{\sin{\varphi_{0}}} + \case{3}{8}
    r_{x0}^{2} + P(\infty) =0.
    \label{rel_3}
\end{equation}

The second regime occurs when $P(\tau)$ tends to infinity with time. In this more general case, almost all nanoparticles with $r_{x0} > 0$ move to the right ($r_{x} \to \infty$ as $\tau \to \infty$) and almost all nanoparticles with $r_{x0} < 0$ move to the left ($r_{x} \to -\infty$ as $\tau \to \infty$) in an oscillatory way. In other words, the harmonically oscillating gradient magnetic field induces the directed transport of suspended nanoparticles if $H_{\perp} \neq 0$. This is at first sight surprising because the static magnetic field $H_{\perp}$ does not generate a magnetic force. However, according to (\textcolor{blue} {\ref{t_d}}), this field generates an additional torque, which affects the magnetization angle $\varphi$ and, hence, indirectly contributes to the driving force (\textcolor{blue} {\ref{f_g}}). It is this torque that is responsible for the appearance of the directed transport of nanoparticles subjected to the harmonically oscillating gradient magnetic field (see below for more details).

In general, the solutions of equations (\textcolor{blue} {\ref{set_1}}) and (\textcolor{blue} {\ref{set_2}}) obtained for arbitrary initial values $\varphi_{0}$ and $r_{x0}$ of the magnetization angle $\varphi$ and particle position $r_{x}$ completely describe the directed transport of suspended nanoparticles. Within the framework of our model, any parameter characterizing this phenomenon can be derived from these solutions. One of the most important parameters is the drift velocity, i.e., the average particle velocity $\overline{v}$, which we define as the particle displacement $\Delta r_{x}$ (after the transient time $\tau_{tr}$) during the period $2\pi$ of the gradient magnetic field: $\overline{v} = \Delta r_{x} / 2\pi$. Since $\Delta r_{x} = r_{x}(2\pi + \tau_{tr}) - r_{x}(\tau_{tr})$, from (\textcolor{blue} {\ref{set_2}}) we find the following representation of the drift velocity:
\begin{equation}
    \overline{v} = \frac{2\nu_{g}}{3\pi}
    \int_{\tau_{tr}}^{2\pi + \tau_{tr}}
    \cos{\varphi (\tau)}\sin{(\tau +
    \phi)} d\tau.
    \label{v}
\end{equation}
Note also that the dimensional drift velocity $\overline{V}$ is expressed through the dimensionless one as $ \overline{V} = \mathrm{\Omega} a\overline{v}$.

\subsection{Drift phenomenon: analytical results}
\label{ssec:4.2}

Since, according to (\textcolor{blue} {\ref{nu}}), $\nu_{\perp}/ \nu_{g} = H_{\perp}/ga$, this ratio is usually sufficiently large. Taking into account the model parameters introduced in the previous section, we restrict our analysis to the case with $\nu_{\perp} \gg 1$ and $\nu_{g} \ll 1$.

\subsubsection{Nanoparticle dynamics near the origin}

We start the study of the drift phenomenon with considering the nanoparticle dynamics in a relatively small vicinity of the origin. More precisely, here we study the rotational and translational motions of nanoparticles, whose initial positions satisfy the condition $1 \lesssim |r_{x0}| \ll \nu_{\perp} / \nu_{g} $. For these nanoparticles there are two very different dynamical regimes. The first occurs at $0 \leq \tau \leq \tau_{tr}$ and is characterized by fast changes of the variables $\varphi$ and $r_{x}$, while the second, which occurs at $\tau \geq \tau_{tr}$, is characterized by their slow dynamics.

The above assumptions allow us to approximately determine the magnetization angle $\varphi$ in these two regi\-mes. Indeed, since $\nu_{\perp} \gg 1$ and, as a consequence, $\tau_{tr} \ll 1$ (this will be shown explicitly later), the functions $\cos{\varphi}$ and $\sin{\varphi}$ on the time interval $[0, \tau_{tr}]$ can be linearly approximated by
\begin{eqnarray}
    &\cos{\varphi} = \cos{\varphi_{0}} +
    (\cos{\varphi^{(0)}} - \cos{\varphi_{0})}
    \tau/ \tau_{tr},
    \nonumber \\[3pt]
    &\sin{\varphi} = \sin{\varphi_{0}} +
    (\sin{\varphi^{(0)}} - \sin{\varphi_{0})}
    \tau/ \tau_{tr}
    \label{cos,sin}
\end{eqnarray}
[$\varphi^{(0)} = \varphi{(\tau_{tr})}$]. Then, by integrating (\textcolor{blue} {\ref{set_1}}) with respect to $\tau$ from 0 to $\tau_{tr}$, one obtains the relation
\begin{eqnarray}
    &2(\varphi^{(0)} - \varphi_{0}) =
    \nu_{\perp} (\cos{\varphi^{(0)}} +
    \cos{\varphi_{0})} \tau_{tr}
\nonumber \\[3pt]
    &- \nu_{g} r_{x0} \sin{\phi}\,
    (\sin{\varphi^{(0)}} +
    \sin{\varphi_{0})} \tau_{tr},
    \label{rel_4}
\end{eqnarray}
which, after calculating $\varphi^{(0)}$, will be used to determine the transient time $\tau_{tr}$.

In the second dynamical regime (when $\tau \geq \tau_{tr}$), we represent the magnetization angle in the form
\begin{equation}
    \varphi = \varphi^{(0)} + \varphi_{1}(\tau),
    \label{varphi_3}
\end{equation}
where $|\varphi_{1}(\tau)| \ll 1$ and $\varphi_{1} (\tau_{tr}) = 0$. If $\tau$ is restricted by a few periods of the gradient magnetic field, then the particle position $r_{x}$ in (\textcolor{blue} {\ref{set_1}}) can be replaced by its initial position $r_{x0}$. With these assumptions, (\textcolor{blue} {\ref{set_1}}) reduces to the linear one
\begin{eqnarray}
    &\dot{\varphi}_{1}(\tau) = \nu_{\perp}
    \cos{\varphi^{(0)}} - \nu_{\perp}
    \sin{\varphi^{(0)}}\varphi_{1}(\tau)
    \nonumber \\[3pt]
    &- \nu_{g} r_{x0}\sin{\varphi^{(0)}}
    \sin{(\tau + \phi)},
    \label{varphi_1_eq}
\end{eqnarray}
whose solution is given by
\begin{eqnarray}
    &\varphi_{1}(\tau) = \cot{\varphi^{(0)}}
    + \frac{\nu_{g} r_{x0}\sin{\varphi^{(0)}}}
    {1 + \nu_{\perp}^{2} \sin^{2}{\varphi^{(0)}}}
    \nonumber \\[3pt]
    &\times \left(\cos{(\tau + \phi)} - \nu_{\perp}
    \sin{\varphi^{(0)}} \sin{(\tau + \phi)}\right).
    \label{varphi_1a}
\end{eqnarray}
Keeping the terms of order $\nu_{g} / \nu_{\perp}$, (\textcolor{blue} {\ref{varphi_1a}}) simplifies to
\begin{equation}
    \varphi_{1}(\tau) = \cot{\varphi^{(0)}}
    - \frac{\nu_{g}}{\nu_{\perp}} r_{x0}
    \sin{(\tau + \phi)}.
    \label{varphi_1b}
\end{equation}
From this, using the condition $\varphi_{1} (\tau_{tr}) = 0$ and representation (\textcolor{blue} {\ref{varphi_3}}), we find for the second dynamical regime
\begin{equation}
    \varphi = \frac{\pi}{2} - \frac
    {\nu_{g}}{\nu_{\perp}} r_{x0}
    \sin{(\tau + \phi)}.
    \label{varphi_4}
\end{equation}

Now we are in a position to calculate the transient time $\tau_{tr}$. Indeed, assuming that the initial magnetization angle $\varphi_{0}$ is not too close to $\pi/2$ and replacing, according to (\textcolor{blue} {\ref{varphi_4}}), $\varphi^{(0)}$ by $\pi/2$ [recall, $\varphi^{(0)} = \varphi(\tau_{tr})$], from (\textcolor{blue} {\ref{rel_4}}) we obtain the explicit formula
\begin{equation}
    \tau_{tr} = \frac{\pi - 2\varphi_{0}}
    {\nu_{\perp} \cos{\varphi_{0}}},
    \label{tau1_1}
\end{equation}
which shows that $\tau_{tr} \ll 1$. If $\varphi_{0} = \pi/2$, then  relation (\textcolor{blue} {\ref{rel_4}}) together with
\begin{equation}
    \varphi^{(0)} = \frac{\pi}{2} -
    \frac{\nu_{g}}{\nu_{\perp}}
    r_{x0} \sin{\phi}
    \label{varphi^0_1}
\end{equation}
(we assume that $\phi \gg \tau_{tr}$) yields $\tau_{tr} = 2/ \nu_{\perp}$. Thus, since from (\textcolor{blue} {\ref{tau1_1}}) it follows that $\tau_{tr} \to 2/\nu_{\perp}$ as $\varphi_{0} \to \pi/2$, we conclude that formula (\textcolor{blue} {\ref{tau1_1}}) can always be used to estimate the transient time.

Finally, using (\textcolor{blue} {\ref{set_2}}) and the above results, it is also possible to describe the translational motion of nanoparticles. Let us first determine the particle displacement $r_{x}^{(0)} - r_{x0}$ [$r_{x}^{(0)} = r_{x}(\tau_{tr})$] during the transient time $\tau_{tr}$. To this end, we substitute $\cos{\varphi}$ from (\textcolor{blue} {\ref{cos,sin}}) into (\textcolor{blue} {\ref{set_2}}) and integrate this equation over $\tau$ from 0 to $\tau_{tr}$. Doing so, and using (\textcolor{blue} {\ref{tau1_1}}) and (\textcolor{blue} {\ref{varphi^0_1}}), for the particle displacement one obtains the formula
\begin{equation}
    r_{x}^{(0)} - r_{x0} = \frac{2\nu_{g}}
    {3\nu_{\perp}}\frac{\pi - 2\varphi_{0}}
    { \cos{\varphi_{0}}} \sin{\phi}\left(
    \frac{\nu_{g}}{\nu_{\perp}} r_{x0}
    \sin{\phi} + \cos{\varphi_{0}}\right)\!.
    \label{r_x^0_1}
\end{equation}
It shows that, depending on the initial conditions, the particle displacements can be very different. In particular, if $|\cos{\varphi_{0}}| \gg (\nu_{g} / \nu_{\perp})|r_{x0}| \sin{\phi}$, then
\begin{equation}
    r_{x}^{(0)} - r_{x0} = \frac{2\nu_{g}}
    {3\nu_{\perp}}(\pi - 2\varphi_{0})
    \sin{\phi},
    \label{r_x^0_2}
\end{equation}
while at $\varphi_{0} = \pi/2$ formula (\textcolor{blue} {\ref{r_x^0_1}}) yields
\begin{equation}
    r_{x}^{(0)} - r_{x0} = \frac{4\nu_{g}^{2}}
    {3\nu_{\perp}^{2}} r_{x0} \sin^{2}{\phi}.
    \label{r_x^0_3}
\end{equation}

To describe the translational motion of nanoparticles in the second dynamical regime, we use (\textcolor{blue} {\ref{set_2}}) with the magnetization angle given in (\textcolor{blue} {\ref{varphi_4}}). By integrating this equation over the time interval $[\tau_{tr}, \tau]$, where, as before, $\tau (\geq \tau_{tr})$ is restricted by a few periods of the gradient magnetic field, we arrive to the following expression for the particle coordinate:
\begin{eqnarray}
    &r_{x} = r_{x}^{(0)} + \frac{\nu_{g}^{2}}
    {3\nu_{\perp}} r_{x0} \Big[ 2(\tau -
    \tau_{tr}) - \sin{2(\tau + \phi)}
    \nonumber \\[3pt]
    & + \sin{2(\tau_{tr} + \phi)} \Big].
    \label{r_x_3}
\end{eqnarray}
According to this result, nanoparticles move in an oscillatory manner to the right (if $r_{x0} > 0$) or to the left (if $r_{x0} < 0$). This confirms our hypothesis that the harmonically oscillating gradient magnetic field generates their directed transport if $H_{\perp} \neq 0$. Using the definition of the drift velocity, $\overline{v} = [r_{x}(2\pi + \tau_{tr}) - r_{x}(\tau_{tr})]/2\pi$, from (\textcolor{blue} {\ref{r_x_3}}) for the drift velocity of nanoparticles near the origin, $\overline{v}_{1} = \overline{v}|_{|r_{x0}| \ll \nu_{\perp} /\nu_{g}}$, one immediately gets
\begin{equation}
    \overline{v}_{1} = \frac{2\nu_{g}^{2}}
    {3\nu_{\perp}} r_{x0}.
    \label{v2}
\end{equation}

The fact that $\overline{v}_{1}$ linearly depends on $r_{x0}$ means that the particle coordinate $r_{x}(2\pi n)$ exponentially grows with the number $n$ ($n=1,2,\ldots$) of periods of the oscillating gradient magnetic field, if the condition $|r_{x}(2\pi n)| \ll \nu_{\perp}/ \nu_{g}$ holds. Indeed, in this case
\begin{equation}
    r_{x}(2\pi n) = r_{x}(2\pi n - 2\pi)
    \bigg( 1 + \frac{4\pi \nu_{g}^{2}}
    {3\nu_{\perp}} \bigg)
    \label{recur}
\end{equation}
and, as a consequence,
\begin{equation}
    r_{x}(2\pi n) = r_{x0} \exp\! \bigg[n
    \ln\! \bigg( 1 + \frac{4\pi \nu_{g}^{2}}
    {3\nu_{\perp}} \bigg) \bigg].
    \label{r_xn}
\end{equation}
To avoid confusion, we stress two points related to this result. First,  since $|r_{x}(2\pi n)| \ll \nu_{\perp}/ \nu_{g}$ and $4\pi \nu_{g}^{2}/3\nu_{\perp} \ll 1$, the particle coordinate $r_{x}(2\pi n)$ grows exponentially with the discrete time $2\pi n$ only if
\begin{equation}
    n \ll n_{\mathrm{max}} = \frac{3\nu_{\perp}}
    {4\pi \nu_{g}^{2}} \ln{\frac{\nu_{\perp}}
    {\nu_{g}|r_{x0}|}}.
    \label{n_max}
\end{equation}
In other words, formula (\textcolor{blue} {\ref{r_xn}}) holds in the time interval $\tau_{tr} < \tau \ll 2\pi n_{\mathrm{max}}$ for nanoparticles with $|r_{x0}| \ll \nu_{\perp} / \nu_{g}$. And second, according to (\textcolor{blue} {\ref{r_x_3}}), the particle coordinate $r_{x}(\tau)$ [in contrast to $r_{x}(2\pi n)$] grows with time $\tau$ in an oscillatory manner.

\subsubsection{Drift velocity far from the origin}

Now we determine the drift velocity of nanoparticles far from the origin, when the condition $|r_{x}| \gg \nu_{\perp} /\nu_{g}$ is satisfied. In this case, the transient time $\tau_{tr} \sim (\nu_{g} |r_{x}|)^{-1}$ is small ($\tau_{tr} \ll \nu_{\perp}^{-1}$) and, according to (\textcolor{blue} {\ref{set_1}}), the magnetization angle $\varphi$ and the particle position $r_{x}$ are connected by the approximate relation
\begin{equation}
    \cot{\varphi} - \frac{\nu_{g} r_{x}}
    {\nu_{\perp}}\sin{(\tau + \phi)} =0.
    \label{rel}
\end{equation}
From this it follows that
\begin{equation}
    \cos{\varphi} = \mathrm{sgn}{(r_{x})}
    \left\{\!\! \begin{array}{ll}
    1, & \;  0< \tau < \pi-\phi,
    \\[3pt]
    -1, & \; \pi-\phi < \tau < 2\pi-\phi,
    \\[3pt]
    1, & \; 2\pi-\phi < \tau < 2\pi
    \end{array}
    \right.
\label{cos_varphi}
\end{equation}
[$\mathrm{sgn}(x)$ is the sign function] as $\nu_{g} |r_{x}| /\nu_{\perp} \to \infty$. Therefore, substituting (\textcolor{blue} {\ref{cos_varphi}}) into (\textcolor{blue} {\ref{v}}) and performing a simple integration, the drift velocity of nanoparticles that are far away from the origin, $\overline{v}_{2} = \overline{v}|_{|r_{x}| \gg \nu_{\perp} /\nu_{g}}$, can be represented in the form
\begin{equation}
    \overline{v}_{2} = \mathrm{sgn}{(r_{x})}
    \frac{8\nu_{g}} {3\pi}.
    \label{v3}
\end{equation}
It should be noted that if $r_{x0}$ is not too close to the origin, then $\mathrm{sgn}{(r_{x})}$ can be replaced by $\mathrm{sgn}{(r_{x0})}$, see  section \textcolor{blue} {\ref{ssec:4.1}}.

Since the drift velocity of such nanoparticles is the same (i.e., it does not depend on $|r_{x0}|$), their coordinates linearly increase (if $r_{x0} > 0$) or decrease (if $r_{x0} < 0$) with the discrete time $2\pi n$. By comparing (\textcolor{blue} {\ref{v3}}) and (\textcolor{blue} {\ref{v2}}), it is seen that $|\overline{v}_{2}| \gg |\overline{v}_{1}|$. Surprisingly, according to (\textcolor{blue} {\ref{v3}}), the maximum of the dimensional drift velocity, $|\overline{V}_{2}| = 4Mga^{2} /9\pi \eta\,$ ($\overline{V}_{2} = \mathrm{\Omega} a \overline{v}_{2}$), which is achieved in the harmonically oscillating gradient magnetic field, is close to the maximum velocity $|V| = 2Mga^{2} /9\eta$ in the case of the time-independent gradient magnetic field \cite{DLP2020}.

\subsection{Drift phenomenon: numerical results}
\label{ssec:4.3}

\begin{figure}
    \centering
    \includegraphics[totalheight=5.5cm]{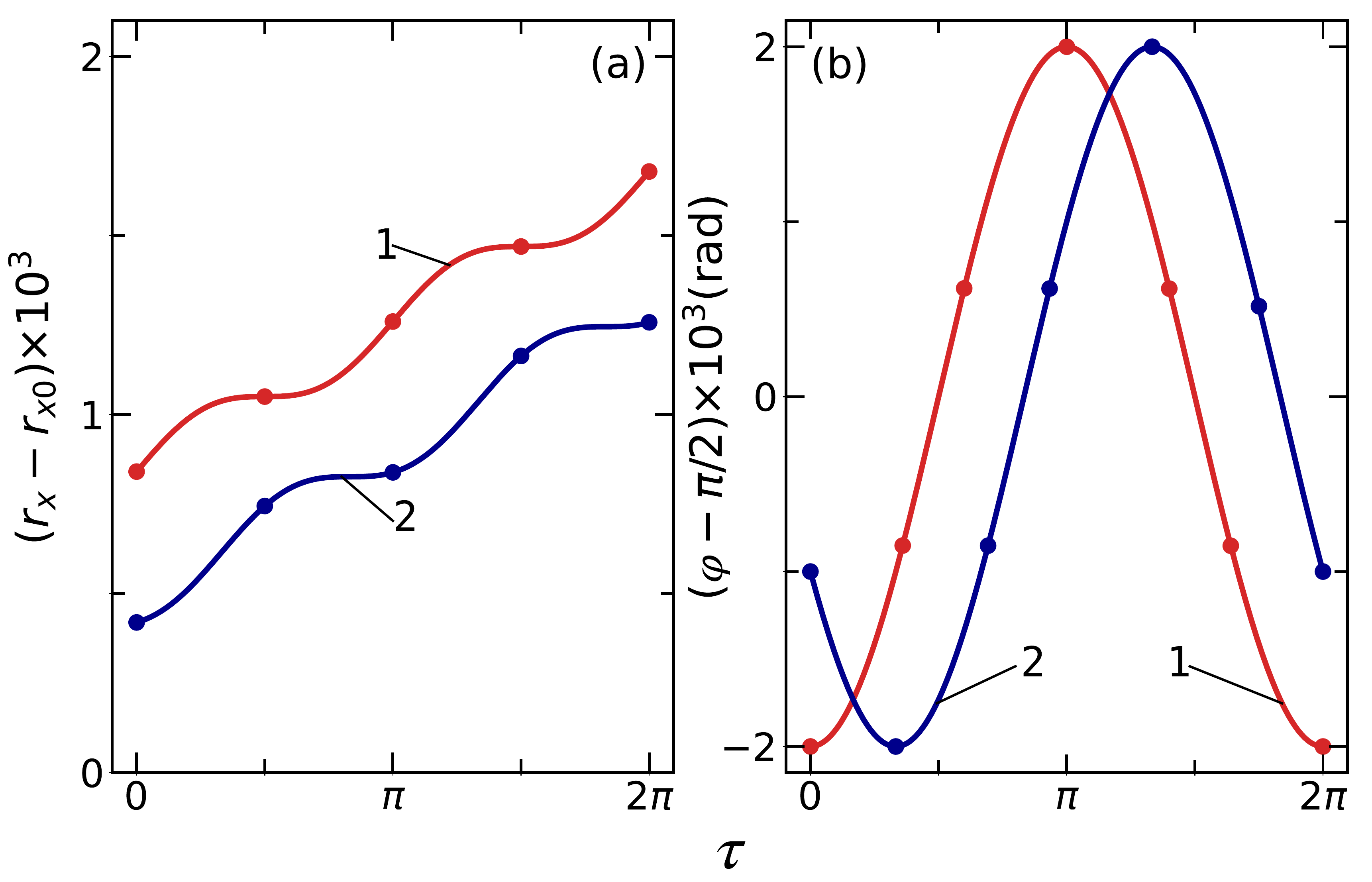}
    \caption{The particle position $r_{x}$ (a) and
    the magnetization angle $\varphi$ (b) as functions
    of the dimensionless time $\tau$ for nanoparticles
    near the origin. The model parameters are chosen
    to be $\nu_{g} = 10^{-1}$, $\nu_{\perp}= 10^{2}$,
    $r_{x0} = 2$ and $\varphi_{0} = 0.3\pi$. The solid
    lines (1) and (2) represent the theoretical results
    (\textcolor{blue} {\ref{r_x_3}}) and (\textcolor{blue} {\ref{varphi_4}}) for $\phi
    = \pi/2$ and $\phi = \pi/6$, respectively.
    The symbols show the results obtained by
    numerical solution of equations (\textcolor{blue} {\ref{set_1}})
    and (\textcolor{blue} {\ref{set_2}}).}
    \label{fig4}
\end{figure}

To validate the theoretical results, we numerically sol\-ved equations (\textcolor{blue} {\ref{set_1}}) and (\textcolor{blue} {\ref{set_2}}) for $\nu_{\perp} \gg 1$. In figure \textcolor{blue} {\ref{fig4}}, we show the theoretical and numerical time evolution of the particle position $r_{x}$ and magnetization angle $\varphi$ for nanoparticles near the origin.
As can be seen, the numerical results (symbols) are in very good agreement with the theoretical predictions. In particular, the functions $r_{x}$ and $\varphi$ in the second dynamical regime are well described by formulas (\textcolor{blue} {\ref{r_x_3}}) and (\textcolor{blue} {\ref{varphi_4}}), and the conditions $r_{x}(2\pi + \tau) = 2\pi \overline{v}_{1} + r_{x}(\tau)$ and $\varphi(2\pi + \tau) = \varphi(\tau)$ hold if $\tau$ is not too large. Moreover, the numerical values of $r_{x}^{(0)}$ and $\varphi^{(0)}$ are in perfect agreement with those obtained from (\textcolor{blue} {\ref{r_x^0_1}}) ($r_{x}^{(0)} - r_{x0} = 8.41 \times 10^{-4}$ for $\phi = \pi/2$ and $r_{x}^{(0)} - r_{x0} = 4.20 \times 10^{-4}$ for $\phi = \pi/6$) and (\textcolor{blue} {\ref{varphi^0_1}}) ($\varphi^{(0)} - \pi/2 = -2 \times 10^{-3}$ for $\phi = \pi/2$ and $\varphi^{(0)} - \pi/2 = -10^{-3}$ for $\phi = \pi/6$). The numerical drift velocity is consistent with the theoretical result (\textcolor{blue} {\ref{v2}}) as well (for the chosen model parameters, $\overline{v}_{1} = 1.33 \times 10^{-4}$).

\begin{figure}
    \centering
    \includegraphics[totalheight=5.5cm]{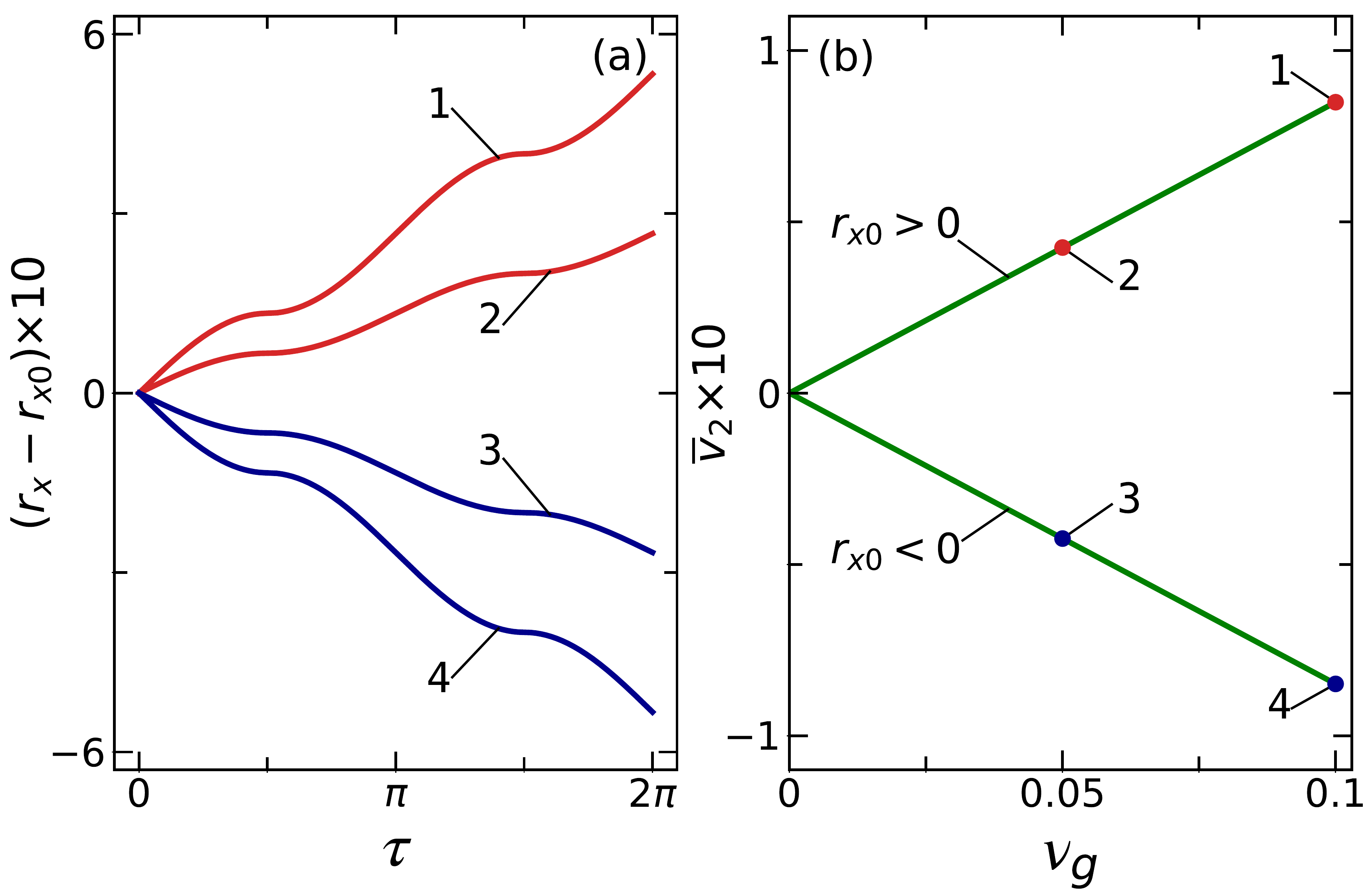}
    \caption{The particle position $r_{x}$ as
    functions of the dimensionless time $\tau$ (a)
    and the limiting average velocity $\overline{v}_{2}$
    as functions of the parameter $\nu_{g}$ (b) for
    nanoparticles with $|r_{x0}| \gg \nu_{\perp} /
    \nu_{g}$. The numerical results for $r_{x}$ are
    shown for $\nu_{g} = 10^{-1}$ and $r_{x0} = 10^{5}$
    (curve 1), $\nu_{g} = 5\times 10^{-2}$ and $r_{x0} =
    10^{5}$ (curve 2), $\nu_{g} = 5\times 10^{-2}$
    and $r_{x0} = -10^{5}$ (curve 3), and $\nu_{g} =
    10^{-1}$ and $r_{x0} = -10^{5}$ (curve 4). The other
    parameters are $\nu_{\perp}= 10^{2}$, $\varphi_{0} =
    \pi/2$ and $\phi = \pi/2$. The theoretical dependencies
    of $\overline{v}_{2}$ on $\nu_{g}$ are calculated
    from (\textcolor{blue} {\ref{v3}}) (they are
    represented by straight lines) and the numerical
    values of $\overline{v}_{2}$ are depicted by symbols.}
    \label{fig5}
\end{figure}

The numerical results that confirm the theoretical formula (\textcolor{blue} {\ref{v3}}) for the average velocity of nanoparticles, whose initial positions are far from the coordinate origin, are illustrated in figure \textcolor{blue} {\ref{fig5}}. In figure \textcolor{blue} {\ref{fig5}b}, the straight solid lines represent the limiting average velocity $\overline{v}_{2}$ as functions of the parameter $\nu_{g}$ for $r_{x0} >0$ and $r_{x0} <0$. The symbols correspond to the velocity $\overline{v}_{2} = [r_{x} - r_{x0}] /2\pi$ determined from the numerical results shown in figure \textcolor{blue} {\ref{fig5}a}. It should be stressed that, since the theoretical result (\textcolor{blue} {\ref{v3}}) is obtained under the condition $|r_{x0}| \gg \nu_{\perp} /\nu_{g}$, the linear dependence of $\overline{v}_{2}$ on $\nu_{g}$ holds if the parameter $\nu_{g}$ is not too small. Note also that the dimensional average velocity of these nanoparticles is rather large: $|\overline{V}_{2}| = 2 \times 10^{3}\, \mathrm{nm\, s^{-1}}$.

\begin{figure}
    \centering
    \includegraphics[totalheight=5.5cm]{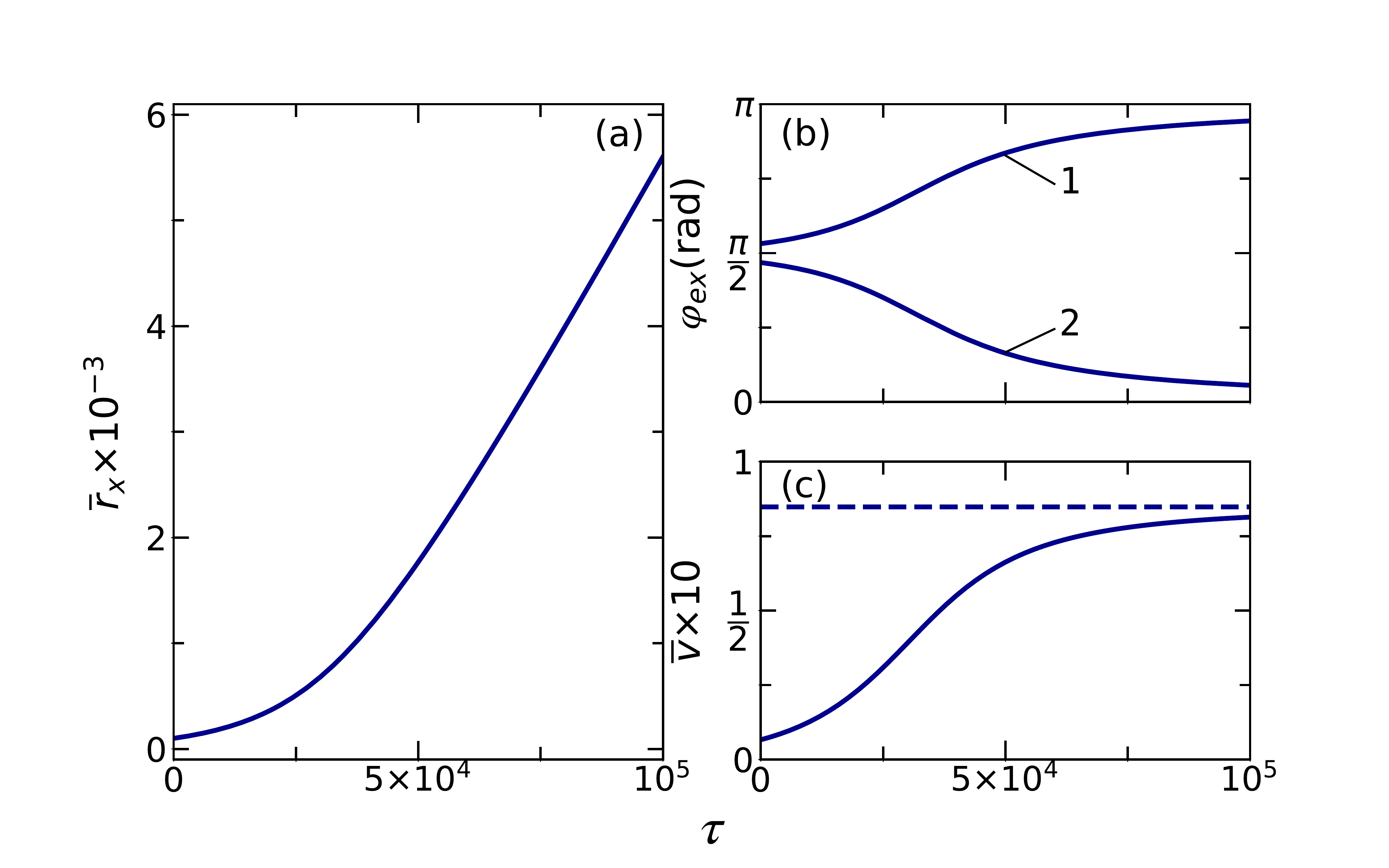}
    \caption{The numerical results for the average
    particle position $\overline{r}_{x}$ (a), the
    extremal values $\varphi_{ex}$ of the magnetization
    angle (b), and the average particle velocity
    $\overline{v}$ (c) as functions of the dimensionless
    time $\tau$. The model parameters are chosen as
    $\nu_{g} = 10^{-1}$, $\nu_{\perp}= 10^{2}$, $r_{x0}
    = 10^{2}$, $\varphi_{0} = \pi/2$ and $\phi = \pi/2$.
    The curves 1 and 2 represent the functions
    $\max{\varphi}$ and $\min{\varphi}$, respectively,
    and the dashed horizontal line corresponds to the
    limiting average particle velocity $\overline{v}_{2}$.}
    \label{fig6}
\end{figure}

Finally, in figure \textcolor{blue} {\ref{fig6}} we show, as an example, the long-time behaviour of the nanoparticle whose initial position $r_{x0}$ equals $10^{2}$. In accordance with our analytical results, the particle moves to the right and the average particle coordinate defined as
\begin{equation}
    \overline{r}_{x} = \frac{1} {2\pi}
    \int_{\tau}^{\tau + 2\pi} r_{x}(\tau')d\tau'
    \label{av_r_x}
\end{equation}
approaches at long times the linear regime, see figure \textcolor{blue} {\ref{fig6}a}. Further, these results show also that the magnetization angle on a short time scale is an almost periodic function of the dimensionless time $\tau$. However, since $|r_{x}|$ grows with $\tau$, one can conclude that $\max{\varphi}$ ($\min{\varphi}$) increases to $\pi$ (decreases to $0$) with increasing $\tau$. This fact is illustrated in figure \textcolor{blue} {\ref{fig6}b}. The time dependence of the average particle velocity $\overline{v} = d \overline{r}_{x} /\tau$ is shown in figure \textcolor{blue} {\ref{fig6}c}. The minimal and maximal values of the average velocity, $\overline{v}_{1} = 6.67 \times 10^{-3}$ and $\overline{v}_{2} = 8.49 \times 10^{-2}$, are in complete agreement with the theoretical ones obtained from (\textcolor{blue} {\ref{v2}}) and (\textcolor{blue} {\ref{v3}}), respectively.
The Python codes used to generate the reported data are deposited to Mendeley Data \cite{MenDat}.

Thus, the results of this section confirm that the drift phenomenon appears only if the uniform magnetic field exists ($H_{\perp}> 0$). This field influences the magnetization angle and, due to this fact, changes the driving force generated by the harmonically oscillating gradient magnetic field. The joint action of these fields results in qualitative changes in the rotational and translational motions of nanoparticles. Specifically, the magnetization angle, which describes the rotational motion, oscillates with time within the interval ($0, \pi$), but the amplitude of oscillations grows slowly with time. The translational motion of nanoparticles also occurs in an oscillatory way, but during each period of the gradient magnetic field they are slightly displaced to the right (if $r_{x0}> 0$) or to the left (if $r_{x0}< 0$). It is these displacements that are responsible for the drift motion of nanoparticles induced by the uniform and time-varying gradient magnetic fields.

\section{Conclusions}
\label{Concl}

We have studied the dynamics of single-domain ferromagnetic nanoparticles in a viscous liquid subjected to both the harmonically oscillating gradient magnetic field and the static uniform magnetic field. Using physically motivated assumptions, we have derived a basic set of first-order differential equations for the magnetization angle and particle coordinate that describe the coupled rotational and translational motions of nanoparticles. To analyse their behaviour, this set of equations has been solved analytically in certain limiting cases and numerically in the general case.

It has been established that, if the uniform magnetic field is absent, the rotational and translational motions of nanoparticles are strictly periodic and, in general, their periods coincide with the period of the gradient magnetic field. We have found analytical expressions for the magnetization angle and particle position for nanoparticles near and far from the coordinate origin, where the gradient magnetic field equals zero. These and other theoretical results have been confirmed by numerical solution of the basic differential equations. An important feature of these equations, which has been taken into account, is that they are stiff.

It has been shown that the presence of even a weak uniform magnetic field qualitatively changes the nano\-particle dynamics making it aperiodic. The most striking observation made is the drift motion of nano\-particles. It occurs in such a way that all nanoparticles with positive (negative) initial positions move to the right (left) with different drift velocities. We have studied this phenomenon by a precise analysis of the basic equations and by analytical and numerical solutions of these equations on short and long time scales. Specifically, for nanoparticles near and far from the origin we have derived  expressions for the magnetization angle, particle position and drift velocity. Since the basic equations of motion are physically meaningful and all our theoretical results are confirmed numerically, it is very likely that the predicted drift phenomenon caused by the harmonically oscillating gradient magnetic field really exists. Its experimental observation would be of great importance for various applications, including drug delivery and separation processes. It seems that this novel mechanism of directed
transport of suspended nanoparticles could also be useful for hyperthermia applications. This is because the harmonically oscillating gradient magnetic field induces not only the nanoparticle transport, but also the nanoparticle heating (due to the rotational and translational oscillations of nanoparticles in a viscous liquid). Since the drift velocity is rather small, the drift motion does not lead to additional heating. However, moving through tumor, such nanoparticles may enhance treatment efficacy.

\ack{This work was partially supported by the Ministry of Education and Science of Ukraine under Grant No.\ 0116U002622. In addition, T.V.L.\ acknowledges the support of the Ukrainian State Fund for Fundamental Research under Grant No.\ F 81/41894.}

S.I.D. and T.V.L. conceptualized the work and conducted theoretical calculations, A.T.L. performed numerical simulations. All authors prepared the manuscript.

\section*{References}
\bibliographystyle{unsrt}

\bibliography{AC_GradField}

\end{document}